\documentclass[twocolumn,prl]{revtex4}

\usepackage{graphics}

\usepackage{epsfig} 

\usepackage{dcolumn}

\usepackage{amsmath}

\begin{document}

\title{Green's Function of Anyons in Calogero Model and Quantum Hydrodynamics}

\author{D. B. Gutman}

\affiliation{Department of Physics, University of Florida,
Gainesville, FL 32611,US}

\date{\today}

\begin{abstract}
We find that correlation functions at one dimension are
crucially affected by the curvature of the bare single particle
spectrum. Parabolic curvature leads to two  closely related
phenomena: the Green's function exhibits oscillation (as a function of
the coordinate), while  the  polarization operator acquires 
support in part of the  frequency-momentum plane. We 
calculated the Green's function using the  WKB approximation
for collective variables theory \cite{JevickiSakita}. 
Within this approach, the single particle
Green's function is related to a quantum soliton \cite{Polychronakos}. 
The finite support of the polarization operator is due to periodic density waves.
\end{abstract}

\maketitle Tomonaga-Luttinger liquid has been a paradigm of non
Fermi liquid systems. The bosonization technique has  been a
particularly efficient way to study it. Recently it  was pointed out
\cite{Pustilnik} that for some one dimensional phenomena the finite
curvature of conductance band is essential. Such are photo-voltaic
effect, thermo-power low temperature Coulomb drag etc. The violation
of particle-hole symmetry is necessary for this and alike effect to
exist. Since conventional bosonization  relies on the linearization
of the single-particle spectrum such phenomena can hardly be
approached.

In this work we address the issue of spectrum curvature for
correlation functions of Calogero-Sutherland (CS)  model\cite{Calogero}.
This model appears in various branches of physics, such as spin chains, FQHE,
disordered metals\cite{reviews,CLP-1979}. The Hamiltonian of CS is
\begin{equation}\label{Hamiltonian}
H=-\frac{\hbar^2}{2m}\sum_{i=1} \partial^2_i +
\left(\frac{\pi}{L}\right)^2\sum_{i>
j}\frac{\lambda(\lambda-1)}{\sin^2\left(\frac{\pi}{L}
(x_i-x_j)\right)}
\end{equation}
The special values of it $\lambda=1,2,4$ corresponds to the Dyson's
symmetry classes of Random Matrix Theory (RMT) \cite{Simons}. The
exact eigenstates and eigenfunctions of CS model have been found,
first by Forrester\cite{Forrester} for integer and later on by
Ha\cite{Ha} for rational values of coupling strength $\lambda$. The
continuous description of correlation function for generic values of
$\lambda$ is absent. A certain progress had been reached in the
Ref.\cite{Gangard} based on the replica symmetry breaking technique.

In this  work  we focus only on long range properties  of CS model.
The structure of this paper is following. Following Awata
\cite{Awata} we pass to the
variables \cite{JevickiSakita}. We explain what part of correlation
functions we are after on the example of free fermions. We calculate
Green's function for the large value of the interaction strength using
saddle-point approximation. Finally we determine the support of the
polarization operator based on the hydrodynamical calculations.

We consider a ring geometry of the system. The coordinates of the
particles can be described by a complex variable
$z_n=Le^{i\theta_n}$ where $\theta$ is an angle along the circle of
radius $L$. The Hamiltonian can be rewritten as
\begin{equation}\label{e1}
H=\left(\frac{2\pi}{L}\right)^2\bigg[\frac{\hbar^2}{2m}\sum_{j=1}^N(z_j\partial_j)^2+\sum_{i \neq j}^N\frac{\lambda(\lambda-1)}{|z_i-z_j|^2}\bigg].
\end{equation}
Ground state of the Hamiltonian (\ref{Hamiltonian}).
$\nolinebreak{\Psi_0=\left(\prod_{i=1}^Nz_i\right)^{-\lambda(N-1)/2}|\Delta|^{\lambda-1}\Delta}$,
$\nolinebreak{\Delta=\prod_{i<j}^N(z_i-z_j)}$.
The excited states
$\nolinebreak{\Psi_{\kappa}=\Psi_0\Phi^B_\kappa}$, where
$\Phi^B_\kappa$ is a symmetric function of the particle coordinates
parameterized by partition $\kappa$. The Hamiltonian $H=\Psi_0^{-1}H
\Psi_0$ is acting  in  the space of symmetric (bosonic) function.
After some algebra one arrives to
\begin{equation}\label{e3}
H=\sum_{i=1}^N D_i^2+\lambda\sum_{i<j}^N\frac{z_i+z_j}{z_i-z_j}(D_i-D_j)
\end{equation}
where $D_i=z_i\partial_i$. To prepare for second quantization one
defines so called collective variables
$\nolinebreak{p(\theta)=\sum_{i=1}^N\delta(\theta-\theta_i), \,\, \,
p_k=\int_0^{2\pi}d\theta e^{ik\theta}p(\theta)}$.
In terms of collective variables the Hamiltonian has a form
\begin{eqnarray}&&
\label{e4} \hspace{-.5cm}H=\!
\frac{1}{2}\!\!\!\sum_{m,n=-N}^Nmnp_{n+m}\frac{\partial^2}{\partial
p_n\partial p_m}\!+\!(1-\!\lambda)\!\sum_{n=-N}^N\!
n^2p_n\frac{\partial}{\partial p_n}\!+\nonumber \\&&
\hspace{-.5cm}\frac{\lambda}{2}\sum_{m=0}^{N-1}\sum_{n=1}^{N-m}(n+m)\bigg[p_np_m\frac{\partial}{\partial
p_{n+m}}+p_{-n}p_{-m}\frac{\partial}{\partial p_{-n-m}}\bigg]
\end{eqnarray}
So far our calculation have been exact. Now we  pass to the
hydrodynamic limit. In the limit of $N \to \infty$ the
Hamiltonian eq.(\ref{e4}) can be written as
\begin{equation}\label{e5}
H=\int dx\bigg[ \frac{1}{2}\rho v^2+U[\rho]\bigg]
\end{equation}
Here  $x$ is a coordinate along the circle
($x=\frac{L}{2\pi}\theta$), the linear density
$\rho(x)=\frac{2\pi}{L}\rho(\theta)$  etc. The potential energy of
quantum liquid is given by
$\nolinebreak{U=\frac{\pi^2\lambda^2}{6}\rho^3-\frac{\pi\lambda(\lambda-1)}{2}\rho^H\rho_x+\frac{(\lambda-1)^2}{8}\frac{(\rho_x)^2}{\rho}}$.
The Hilbert transform is defined as $\nolinebreak{
\rho^H(x)=\frac{1}{\pi}P \int dx' \frac{\rho(x')}{x'-x}}$. The modes
of velocity operator are defined as
$v_n=2\pi\left(-n\frac{\partial}{\partial p_{-n}}+\frac{1}{2}p_n{\rm
sgn(n)}\right)$. It is easy to see that with these definitions
particle velocity and density satisfy a standard commutation
relations one expect for any hydrodynamic theory \cite{Landau}
$[v(x),\rho(y)]=-i\delta'(x-y) $. To reproduce the conventional
bosonization theory one expands the Hamiltonian up to second order
in fluctuations (due to conservation laws the velocity $v$ and
density operators $\delta \rho$ are  of the same order) and spatial
gradient (keeping only the terms that have no gradient). By doing so
we see that on the conventional Luttinger model interaction enters
only through sound velocity of bosonic modes.

Now we use the Hamiltonian (\ref{e5}) to study effects of the
spectrum curvature. We start with single particle Green's function
$G(x,t)=-i\langle T\Psi_A(x,t)\Psi_A^\dagger(0,0)\rangle$,
where $\Psi$ is an anyonic (for $\lambda \neq 1$)
field. Interesting on its own (it's spectrum function can
be measured in angle resolved tunneling experiments
\cite{Auslaender}) Green's function will also be helpful
for study of more involved correlators.
As we will see shortly, the Green's function for CS model
and for free fermions turn out somewhat similar. We consider non interacting case first. By the Wick theorem the Green's function can be
easily evaluated
For parabolic dispersion  the time ordered Green's function ($t>0$) is
given by
\begin{eqnarray}&&
\label{e8} \hspace{-0.5cm}G=-\sqrt{\frac{\pi m}{2t}}e^{i\pi/4}\bigg[
e^{ip_F x\!+\!\frac{im}{2t}(x\!+\!vt)^2}
\rm{Erfc}\left(e^{i\pi/4}\!\sqrt{\frac{m}{2t}}(x\!+\!vt)^2\right)\nonumber
\\&& + e^{-ip_F x+\frac{i
m}{2t}(x-vt)^2}\rm{Erfc}\left(-e^{i\pi/4}\sqrt{\frac{m}{2t}}(x-vt)^2\right)
\bigg]
\end{eqnarray}
The Green's function is a sum of left and right moving terms,
multiplied by fast oscillating term. The smooth part of the Green's
function for right moving particles is plotted on Fig(1). 
The left tail of Green's function is captured by linear bosonization.
There is no right tail, because we deal with strictly parabolic spectrum. The
oscillations  are  universal. The right tail will appear as soon as
negative spectrum curvature is take into account, to match a linear
bosonization.   A transition between oscillating a power law region
is model dependent.

One can get a good idea about function using a saddle point approximation
for the integral over momentum.
\begin{eqnarray}&&
\label{e8} \hspace{-0.5cm}G=\!-i\sqrt{\frac{\pi m
}{t}}\!\bigg[e^{ip_F x}
e^{i(\!x\!-\!vt)^2/2t}\theta(\!x\!-\!vt)\!+\!(L\leftrightarrow R)
\bigg] \nonumber
\end{eqnarray}
One sees that  saddle point gives a dominant contribution only if
the particle moves faster than the Fermi velocity. In this region
the result is oscillatory function. For the particles moving slower
than Fermi velocity the saddle point lies out of the region of
integration and the integral is determined by its low limit. In that
case we restore a standard power decay of the Green's function. The
result for particles moving vaster than Fermi velocity particle does
not agree with linear bosonization prediction
$G(x,t)=-\frac{1}{2\pi}\bigg[\frac{e^{-ip_F x}}{v_Ft+x}+
(R\leftrightarrow L)
\bigg]$. The reason of  this
disagreement is simple. We have evaluated the Green's function of free
fermion with a parabolic spectrum. In metals electrons are subject
to periodic potential. They form Bloch states and have periodic
spectrum that has a negative curvature close to the boundary of
Brillouin zone. This limits the maximally accessible velocity of the
classical trajectory. The  standard bosonization considers only the
situation in which there is no classical trajectory available.

Indeed, within a tight-binding model the single particle spectrum is given by
$\epsilon(p)=W\left(1-\cos\left(\frac{pa}{2\hbar}\right)\right)$,
where $a$ is a lattice constant and $W$ is a band width. We see that
integral has a stationary point at
$\frac{2\hbar}{a W} \frac{x}{t}=\sin\left(\frac{pa}{2\hbar}\right)$
The right hand side of this equation is  a periodic function of
momentum and it is smaller than one. So the saddle point
contribution disappears if the velocity on the classical trajectory
exceeds the one available by band structure ($\hbar aW \ll (x-v_Ft)/t$). This is a
linear bosonization limit, intensively studied in the literature so
far \cite{Stone-book,Tsvelik}.
We will consider the situation when the classical trajectory dominates the result,
but the distance to the light cone is still large
($v_F \ll (x-v_Ft)/t \ll \hbar aW$). In this case the
semiclassical approximation is justified. Close to the light cone
($(x-v_Ft)/t \ll v_F$) WKB approximation fails and other methods are
needed. In all cases the absolute values of times and coordinates,
measured in the Fermi energy and wave length, are large.
\begin{figure}[b]
\label{fig1}
\includegraphics[width = 0.4\textwidth]{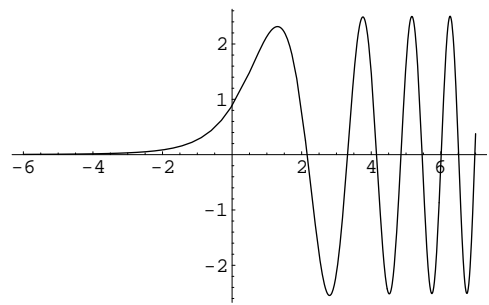}
\caption{single particle Green's function (of right moving electrons)
as a function of $x$ for the finite band-width, $v=1,m=1,t=1$}
\end{figure}

The rest of this paper is devoted to study of asymptotic properties
of Calogero model in the semiclassical regime. We consider the Green's
function first. Anyon  operator is represented via boson
operators\cite{Ha} as
\begin{equation}\label{e11}
\Psi_A(x,t) \simeq \exp\left(i\int_{-\infty}^x
dx(v(x,t)+\pi\lambda\rho(x,t)\right)
\end{equation}
This definition is consistent\cite{BCH} with the commutation relations
\begin{equation}\label{e11a}
\Psi_A(x)\Psi_A(x')= e^{-i\pi\lambda{\rm sgn}(x'-x)}\Psi_A(x')\Psi_A(x).
\end{equation}

The Green's function is given by a functional integral
\begin{equation}\label{e11b}
G_A(x',t')=\frac{-i}{Z}\int\!{\cal{D}}\rho\!{\cal{D}}v
e^{iS}\Psi_A(x',t')\Psi^\dagger_A(0,0) \,\, .
\end{equation}
The hydrodynamic action corresponding to the Hamiltonian eq.(\ref{e5}) is
given by
\begin{equation}
\label{action}
S=\int dxdt\bigg[-v\partial_x^{-1}\partial_t\rho-\frac{1}{2}\rho v^2-U[\rho]\bigg]
\end{equation}
To calculate the Green's function we use steepest decent method, known
also as WKB. The action (\ref{action}) has massive direction and low
energy directions in the configuration space (a constant of shift
$x\to x+a,t\to t+b$) does not change the value of the action on any
solution). The first may be integrated out by saddle point method,
the latter need to be taken into account exactly.
The  saddle point equation is
\begin{equation}\label{e11d}
\rho_t+\partial_x(\rho v)=-\delta(x-x')\delta(t-t')+\delta(x)\delta(t)
\end{equation}
\begin{equation}\label{e12}
v_t+vv_x+\partial_x\left(\frac{\delta U}{\delta \rho}\right)=
-\pi\lambda\delta(x-x')\delta(t-t')+\pi\lambda\delta(x)\delta(t)
\end{equation}
As was suggested earlier \cite{Polychronakos,Abanov} the elementary
particle of fermionic theory corresponds to the quantum soliton of
the boson  theory \cite{Rajaraman}. In the limit of large $\lambda$
we look for solution ($0<t<t'$) in a form
\begin{eqnarray}\label{e13}
\hspace{-0.5cm}\rho(x,t)\!=\!1\!+\!\rho_s(x\!-\!vt\!-\!X(t)),
v(x,t)\!=\!v_s(x\!-\!vt\!-\!X(t)),
\end{eqnarray}
where $X(t)$ is a global variable. It corresponds to the motion of the soliton as a whole.
This is a zero mode of the system.
The soliton's density and velocity are given by \cite{Polychronakos}
\begin{eqnarray}\label{e15}
\rho_s(x)=\frac{B}{\pi}\frac{1}{x^2+B^2}\,\,\, ,\,\,
v_s(x)=\frac{Bv}{\pi}\frac{1}{x^2+B^2+\frac{B}{\pi}}.
\end{eqnarray}
Here $B=\pi\lambda^2/(v^2-\pi^2\lambda^2)$, $v=x'/t'$.  The soliton
exists, provided  $B>0$, i.e.  it's  velocity  {\it exceeds} the
sound velocity $s=\pi\lambda\rho_0$. It is very similar to the
saddle point contribution for the ordinary integral of free fermions
where saddle point contribution appeared only for supersonic
trajectories. As velocity increases $v \gg s$ the size of soliton
$B$ goes to zero and the particle is more localized. By integrating
over infinitesimal region around $x=0,x=x'$ over $x$ we see that the
solution (\ref{e15}) indeed has a right singularity. The total
number of electrons carried by the soliton ($\int \rho(x)dx=1$)  is
quantized to one. The velocity of soliton is  not quantized. Then we
exponential accuracy we find  that Green's function of anyon excitation 
in  CS model is given by
\begin{eqnarray}
G_A(x',t')\simeq e^{iS_{cl}} \int_{X(0)=0}^{X(t')=0} {\cal{D}}{X}e^{\frac{i
m}{2}\int_0^{t'} dt \dot{X}^2(t)}.
\end{eqnarray}
Here the value of the action on classical trajectory
 $S_{cl}=t'(v^2/2-sv)=\frac{x'^2}{2t'}-sx'$.
Performing the integration over zero mode  
find
\begin{eqnarray}&&\label{e18a}
\hspace{-1cm}G_A(x,t)\!\simeq\!\sqrt{\frac{m}{t}}\!\bigg[e^{ip_F\lambda
x\!+\!i(x\!-\!st)^2m/2t}\theta(x\!-\!st)\!+\!(R\!\leftrightarrow\!L)\bigg]
\end{eqnarray}
where   $p_F=\pi\rho_0$ is Fermi momentum. 
Equation (\ref{e18a}) is the central result of this work.
It shows that far from light cone 
the anyon's propagator  oscillates as a function of $x$
with the period $\sqrt{t}$, c.f.  the Green's function of free
electrons eq. (\ref{e8}).

The soliton that determines the Green's function is also responsible
for the finite support of polarization operator
\cite{Polarization_operator_for_free_fermions}
$\Pi (q,t)=-i\langle T\rho(q,t)\rho(-q,0)\rangle$
This, as it is was pointed out in Ref. \cite{Pustilnik}, is
necessary for various physical phenomena. Several attempts had been
in the literature to account for it \cite{Samokhin,Pirooznia}. Here
we show how broadening of the support can be understood within
quantum hydrodynamics.

The polarization operator is determined by neutral particle hole
excitations. In lineal bosonization they are acoustic phonons. In
its nonlinear version they are periodic density
\cite{Polychronakos,Matsuno-book}.
The saddle point equation on periodic waves reduces to
\begin{equation}\label{e23}
\frac{v^2}{\lambda^2}\left(\frac{1}{\rho^2}-1\right)+\pi^2\rho^2+2\pi\rho_x^H-\frac{1}{2}\frac{\rho_{xx}}{\rho}+\frac{1}{4}\frac{\rho_x^2}{\rho^2}=C,
\end{equation}
where $C$ is an arbitrary constant. Solving this equation
\cite{Polychronakos} one finds
\begin{equation}
\label{e24}\rho_{B,l}(x)=1-\frac{1}{l}+\frac{1}{l}\frac{\sinh\left(\frac{2B}{l}\right)}{
\cosh\left(\frac{2B}{l}\right)-\cos\left(\frac{2\pi x}{l}\right)}.
\end{equation}
Here  velocity of the wave
\begin{equation}
\label{e25} v=s\left(1-\frac{1}{l}\right)\sqrt{1+
\frac{4}{l}\left(1-\frac{1}{l}\right)n_B\left(\frac{2B}{l}\right)},
\end{equation}
where $n_B$ is Bose distribution function. In a momentum
representation
\begin{equation}
\label{e25a} \rho_{B,l}(q)=e^{-B|q|/\pi} \sum_n\delta(ql-2\pi n).
\end{equation}
From eq. (\ref{e25},\ref{e25a}) we find that in a long length limit
\begin{equation}
\label{a25b}
q=\frac{2\pi n}{l}\, , \,\,\, v\simeq
s+\frac{2\pi\lambda}{l}\bigg[n_B-\frac{1}{2}\bigg]
\end{equation}

If the width of the picks is bigger than the distance between them
($B \gg l$) one recovers a dispersion of  anharmonic acooustic
phonons $v(q)=s-\lambda q/2m$. This excitations  determine the lower
edge of polarization operator support. In standard bosonization only
those are taken into account
\cite{Polarization_operator_in_linear_bosonization}.

For the waves with period longer than $l \gg B$ there is only singe
pick (soliton) in that length.  One can assume that since other pick
are far apart one can study its motion regardless of others. It
contributes to the energy $E=\frac{\rho_0v^2}{2}$. By recalling
\cite{soliton} the soliton dispersion ($\epsilon(q)=sq+q^2/2m, \,\,
q>0$) we obtain an upper support of polarization operator.

A bosonic and a fermionic descriptions are dual \cite{Stone-book}. In
fermionic  language excitations are characterized by partition
$\kappa$. In bosonic notations  a single period 
neutral excitations are described by the
parameters $B$ and $l$, or alternatively by $n$ and $n_B$. The
partition $\kappa=\{q\}$ corresponds to taking the topmost electrons
and moving it $q$ steps up and corresponds to the diluted gas of
solitons. The conjugated partition $\kappa=1^q$ that describes
moving the top $q$ electrons by one step up corresponds to the
anharmonic acoustic phonons.

To summarize:
the saddle point trajectory that corresponds to the 
soliton solution of non linear hydrodynamic equation 
is dominant for anyon Green's  function. 
By evaluation the value of the action on this trajectory we have 
found an explicit formula for the Green's function of anyon. 

Due to the infinite number of periodic density 
waves a polarization operator acquires a finite
support in frequency  momentum plane. Of which the lower edge of
excitations is given by dispersive acoustic phonons (such excitation
can be accounted  by linearizing an action but accounting phonon
dispersion even for  generic  interaction). The upper edge is
determined by a superposition of non interacting solitons.  As the
model became non integrable and soliton is destroyed the upper part
of supports is expected to smear.

Note added: As this work was  nearly completed such investigation had
been performed in Ref. \cite{Khodas}. It was indeed found that for
generic interaction the upper boundary of the support becomes
smeared, while the lower boundary is unchanged.

I benefited a lot the interaction with Ilya Gruzberg and Paul Wiegmann.
I am grateful to A. Abanov, A. Kamenev, 
D. Maslov, A. Mirlin, M. Stepanov, M. Stone, R. Teodorescu
for useful discussions.






%


\begin{thebibliography}{99}
\bibitem{Polarization_operator_for_free_fermions}
for free electrons the polarization operator is given by
\begin{equation}\label{e20}
\Pi(\omega,q)=\frac{1}{\pi q} \ln \left(
\frac{(\omega+q+\frac{q^2}{2})(\omega-q-\frac{q^2}{2})}
{(\omega-q+\frac{q^2}{2})(\omega+q-\frac{q^2}{2})} \right)
\end{equation}
\bibitem{Polarization_operator_in_linear_bosonization}
Polarization operator within lineral bosonization is
\begin{equation}
\Pi_-(q,\omega)=\frac{2\omega(q)S(q)} {(\omega-\omega(q)+i0{\rm sgn
}  \omega) (\omega+\omega(q)+i0{\rm sgn } \omega)},
\end{equation}
It has a line $\omega=\omega(q)$ as a support on $\omega,q$ plane.
\bibitem{soliton}
There is a difference between the soliton that was discussed in the context of Green's function
and corresponds to the particle added to the system while here it represent
the particle excited from the Fermi surface.
We corrected the energy of the former by hand since the particle conservation
is beyond the hydrodynamics picture.
\end{thebibliography}
\end{document}